# Freedom of expression and 'right to be forgotten' cases in the Netherlands after Google Spain

Stefan Kulk & Frederik Zuiderveen Borgesius

Contact: Frederikzb[at]cs.ru.nl

*Abstract*

*Since the Google Spain judgment of the Court of Justice of the European Union, Europeans have, under certain conditions, the right to have search results for their name delisted. This paper examines how the Google Spain judgment has been applied in the Netherlands. Since the Google Spain judgment, Dutch courts have decided on two cases regarding delisting requests. In both cases, the Dutch courts considered freedom of expression aspects of delisting more thoroughly than the Court of Justice. However, the effect of the Google Spain judgment on freedom of expression is difficult to assess, as search engine operators decide about most delisting requests without disclosing much about their decisions.*

**Table of Contents**

## I    Introduction

Since the *Google Spain* judgment of the Court of Justice of the European Union,[1] Europeans have, under certain conditions, the right to have search results for their name removed from the search results (hereafter: delisted). In this article we examine how the *Google Spain* judgment is applied in the Netherlands, in particular how the right to freedom of expression was valued.[2]

Section 2 explains that several fundamental rights are issue when delisting search results is concerned: the right to privacy and data protection, and the right to freedom of expression. This section also summarises the main points of the *Google Spain* judgment.

---

[1] Case C-131/12 *Google Spain SL and Google Inc. v Agencia Española de Protección de Datos (AEPD) and Mario Costeja González* [2014] not yet published. For a case note see: H Kranenborg, 'Google and the Right to Be Forgotten', (2015) 1(1) EDPL 70.

[2] Especially in section II and IV.1 we build on our earlier work, and borrow some phrases from our earlier work. See: S Kulk and FJ Zuiderveen Borgesius, 'Google Spain v. González: Did the Court Forget About Freedom of Expression?' (2014) 5(3) EJRR 389. We thank Egbert Dommering, Ronan Fahan, Dirk Henderickx, Judith Rauhofer, Mistale Taylor, Rachel Wouda, and the anonymous reviewers for comments on drafts of this paper. Any errors are the authors' own.

Section 3 discusses how the Dutch Data Protection Authority and Dutch courts have dealt with delisting requests since *Google Spain*. In section 4 we comment on the Dutch practices so far.

In section 5 we conclude: search engine operators do not offer much transparency about how they decide on delisting requests. Therefore, the effects of *Google Spain* on freedom of expression in the Netherlands cannot be fully assessed. However, in the few court cases on delisting in the Netherlands, courts paid more attention to freedom of expression than the CJEU did in *Google Spain*.

## II    Delisting search results and fundamental rights

### II.1    Privacy, data protection, and freedom of expression rights

On the internet, an immense amount of information is available, including information about people. Some online information is short-lived;[3] other online information remains available for a long time.

Search engines enable people to find online information, including information about themselves or other people. Without a search engine, online information about people would be laborious to find. In addition, when presented as a search result, the original context of information is not always evident. Search engines re-contextualise the original content by presenting excerpts of it in the search results list, alongside information from other sources that is relevant to the search query. Relevancy is determined by the search engine through an opaque ranking process, over which search engine users can exercise only little control.[4] If you search for a person's name, a search engine presents you with a list of information from different sources. That set of information, as a whole, can create an image of that person. If one had to sift through all available information online without the help of a search engine, such an image would be difficult to construe.[5]

---

[3] See in the legal context: J Zittrain, K Albert and L Lessig, 'Perma: Scoping and Addressing the Problem of Link and Reference Rot in Legal Citations' (2013) 127 Harvard Law Review Forum, 176: "49.9% of the links cited in the [US] Supreme Court opinions no longer had the cited material."

[4] F Pasquale, *The Black Box Society. The Secret Algorithms that Control Money and Information* (Harvard University Press 2015) 59.

Hence, not only the publication of information relating to people online is relevant to people's privacy, but also the search engine's activities regarding that information. People who want to have search results delisted can invoke the right to privacy and the right to protection of personal data.[6] These rights are enshrined in Article 7 and 8 of the Charter of Fundamental Rights of the European Union.

When search results are delisted, several fundamental rights are at issue – not only privacy and data protection rights, but also freedom of expression.[7] None of these rights is absolute. This makes questions about delisting search results difficult, as a proper balance must be struck between the conflicting rights.

The right to freedom to receive and impart information, the right to freedom of expression for short, is affected when search results are delisted. The Charter of Fundamental Rights of the European Union and the European Convention on Human Rights both protect the right 'to receive and impart information and ideas without interference by public authority and regardless of frontiers.'[8]

When information is delisted, at least three types of parties can claim their right to freedom of expression. First, people who publish information on the web have a right to impart information. The European Court of Human Rights notes that the right to freedom of expression protects not only the expression (such as a publication) itself, but also the means of communicating that expression: 'Article 10 [of the European Convention on Human Rights] applies not only to the content of information but also to the means of transmission or reception since any restriction imposed on the means necessarily interferes with the right to receive and impart information.'[9] Hence, if people publish information on the web and their publications are harder to find due to a delisting, their freedom to impart information is interfered with.[10]

---

[5] I Ruthven, C Clews and WHM Dali, First impressions: how search engine results contextualise digital identities' (2010) Proceedings of the third symposium on Information interaction in context, 314: "We argue that the results of an aggregated search can provide a context within we which we can form initial judgments about a person."
[6] We use the "right to privacy" and "the right to respect for private life" interchangeably in this paper. See on the distinction González Fuster G, *The Emergence of Personal Data Protection as a Fundamental Right of the EU* (Springer 2014), 255.
[7] A discussion of the freedom to conduct a business, protected by art 16 of the Charter of Fundamental Rights of the European Union, falls outside the scope of this paper.
[8] Art 11(1) of the Charter of Fundamental Rights of the European Union and art 10 of the European Convention on Human Rights.

Second, as the Advocate General in the *Google Spain* case remarked, a "search engine service provider lawfully exercises (…) his (…) freedom of expression when he makes available internet information location tools relying on a search engine."[11] Indeed, an organised list of search results could be considered a form of expression.[12] Moreover, as Van Hoboken notes, in principle search engine operators can invoke their right to receive information for crawling and indexing web pages.[13] Furthermore, search engine operators should have a right to impart information regarding how they present search results.

Third, searchers have the right to receive information. As the European Court of Human Rights notes, 'the public has a right to receive information of general interest.'[14] Furthermore, 'the internet plays an important role in enhancing the public's access to news and facilitating the sharing and dissemination of information generally (…)'.[15] The public finds that information through search engines. Hence, search engines are means to realise people's right to receive information.

However, the right to freedom of expression is not absolute, and must be balanced against other rights, such as privacy and data protection rights. The European Court of

---

[9] *Autronic AG v Switzerland* App no 12726/87 (ECtHR, 22 May 1990) [47].
[10] JVJ van Hoboken, *Search Engine Freedom. On the Implications of the Right to Freedom of Expression for the Legal Governance of Web Search Engines* (Kluwer Law International 2012) 350.
[11] Case C-131/12 *Google Spain SL and Google Inc. v Agencia Española de Protección de Datos (AEPD) and Mario Costeja González* [2014]. Opinion AG Jääskinen [132].
[12] In the United States, some judges have granted search engines such freedom of expression claims (on the basis of the First Amendment of the US Constitution). E.g. *Search King, Inc. v. Google Technology, Inc.*, 2003 WL 21464568 (W.D. Okla. 2003). For a discussion see: E Volokh and DM Falk, 'Google First Amendment Protection for Search Engine Search Results' (2011-2012) 82 Journal of Law, Economics and Policy 883. For criticism on granting such claims, see: O Bracha, 'The Folklore of Informationalism: The Case of Search Engine Speech' (2014) 82 Fordham Law Review 1629.
[13] Van Hoboken (supra, note 10) 351. Website publishers can prevent search engines from indexing their page by including a 'nofollow' tag or a 'robots.txt' file.
[14] *Társaság a Szabadságjogokért v Hungary* App no 37374/05 (ECtHR 14 April 2009) [26].
[15] *Fredrik Neij and Peter Sunde Kolmisoppi v Sweden* App no 40397/12 (ECtHR 19 February 2013) (inadmissible).

Human Rights says in the context of press publications about freedom of expression and privacy: 'as a matter of principle these rights deserve equal respect.'[16] To strike that balance, the European Court of Human Rights has developed a set of criteria. In a recent case, that Court summarises them as follows:

> (i) contribution to a debate of general interest;
> (ii) how well-known is the person concerned and what is the subject of the report;
> (iii) prior conduct of the person concerned;
> (iv) method of obtaining the information and its veracity (…);
> (v) content, form and consequences of the publication; and
> (vi) severity of the sanction imposed [on the party claiming an interference with freedom of expression].[17]

## II.2    The CJEU's Google Spain judgment

The Google Spain judgment of the CJEU was triggered by a Spanish dispute between Mr. Costeja Gonzáles and Google. If people searched Google for the name of Costeja Gonzáles, Google linked to a tiny newspaper announcement from 1998, concerning a real estate auction to recover social security debts of Mr. Costeja Gonzáles.[18] Without Google's search engine, the newspaper announcement would probably have faded from memory, hidden by practical obscurity.[19] Costeja Gonzáles wanted Google to delist the search results, because the announcement suggesting he had financial problems was out-dated. He complained to the Spanish Data Protection Authority; the case eventually made it to the CJEU.

---

[16] *Axel Springer AG v Germany* App no 39954/08 (ECtHR 7 February 2012) [87]. See similarly: *Von Hannover v Germany* App nrs 40660/08 and 60641/08 (ECtHR 7 February 2012) [100]; *Węgrzynowski and Smolczewski v Poland* App no 33846/07 (ECtHR 16 July 2013) [56].
[17] *Satakunnan Markkinapörssi Oy And Satamedia Oy v. Finland* App no 931/13 (ECtHR 21 July 2015) [62]. The ECtHR refers to *Axel Springer AG v Germany* App no 39954/08 (ECtHR 7 February 2012) and *Von Hannover v Germany* App nrs 40660/08 and 60641/08 (ECtHR 7 February 2012).
[18] See for the original publication: <http://hemeroteca.lavanguardia.com/preview/1998/01/19/pagina-23/33842001/pdf.html> last accessed on 6 August 2015.
[19] We borrow the 'practical obscurity' phrase from the US Supreme Court: Dep't of Justice v. Reporters Comm. for Freedom of the Press, 489 U.S. 749, 762 (1989). For an analysis see: KH Youm and A Park, '"The Digital Right to be Forgotten" in EU Law: Informational Privacy vs. Freedom of Expression', paper presented at the annual meeting of the Association for Education in Journalism and Mass Communication, San Francisco, CA, 6 August 2015.

The CJEU asserted in its *Google Spain* judgment that people have, under certain circumstances, the right to have search results for their name delisted. This right can extend to lawfully published information, such as the information about Costeja Gonzáles.

Articles 12(b) and 14(a) of the EU Data Protection Directive grant data subjects the right to request erasure of personal data, and the right to object to processing personal data.[20] The right to have search results delisted is based by the CJEU on the Data Protection Directive and the privacy and data protection rights of the Charter.[21]

According to the CJEU, a search engine operator processes personal data if it indexes, stores, and refers to personal data available on the web.[22] The CJEU sees the search engine operator as a 'data controller' in respect of this processing, which implies that the operator must comply with data protection law.[23] The CJEU adds, without explanation, that a search engine operator cannot rely on the exception in data protection law for data processing for journalistic purposes.[24] The English version of the judgment says Google 'does not appear' to be able to benefit from the media exception.[25] However, in the authentic language of the judgment, Spanish, the CJEU says Google cannot benefit from the media exception.[26]

---

[20] European Parliament Council Directive 95/46/EC of 24 October 1995 on the protection of individuals with regard to the processing of personal data and on the free movement of such data [1995] OJ L281/31.
[21] *Google Spain* (supra, note 1) [99].
[22] *Google Spain* (supra, note 1) [28].
[23] *Google Spain* (supra, note 1) [33].
[24] In the *Satamedia* case, the CJEU has interpreted the media exemption very broadly (Case C-73/07, *Tietosuojavaltuutettu v. Satakunnan Markkinapörssi Oy and Satamedia Oy* [2008] ECR I-09831, dictum).
[25] *Google Spain* (supra, note 1) [85]. The English (and the French) version say Google 'does not appear' to be able to benefit from the media exception, and thus incorrectly imply that Google might benefit from the exception. See on the media exception: D. Erdos, 'From the Scylla of Restriction to the Charybdis of Licence? Exploring the scope of the "special purposes" freedom of expression shield in European data protection', Common Market Law Review, Vol. 52 (1), 119-153 (2015).
[26] See Art. 41 of the Rules of Procedure of the Court of Justice of the European Union. In Spanish, the CJEU says Google cannot benefit from the media exception: '*ése no es el caso en el supuesto del*

The CJEU says a search engine can provide searchers a 'detailed profile' of a data subject, thereby 'significantly' affecting privacy and data protection rights.[27] Search results for a name provide 'a structured overview of the information relating to that individual that can be found on the internet – information which potentially concerns a vast number of aspects of his private life and which, without the search engine, could not have been interconnected or could have been only with great difficulty.'[28]

Every data subject has the right to correct or remove personal data that are not processed in conformity with the Data Protection Directive.[29] Not only inaccurate data can lead to such unconformity, but also data that are inadequate, irrelevant or excessive in relation to the processing purposes, for instance because they have been stored longer than necessary.[30] Therefore, says the CJEU, the data subject has a right to demand delisting of search results for his or her name.

The CJEU says 'a fair balance' must be struck between the legitimate interests of searchers and the privacy and data protection rights of the data subject.[31] The CJEU adds that the data subject's privacy and data protection rights override, 'as a rule', the search engine operator's economic interests, and the public's interest in finding information.[32] However, the CJEU stresses that data subjects' rights should not prevail if the interference with their rights can be justified by the public's interest in accessing information, for example, because of the role played by the person in public life.

The *Google Spain* judgment has been criticised for its limited consideration of the right to freedom of expression.[33] We have noted elsewhere that the CJEU does not

---

*tratamiento que lleva a cabo el gestor de un motor de búsqueda*' ('this is not the case in the event of the processing [of personal data] that the operator of a search engine carries out').
[27] *Google Spain* (supra, note 1) [38]; [80].
[28] *Google Spain* (supra, note 1) [80].
[29] Data Protection Directive, art 12(b) and 14(a).
[30] *Google Spain* (supra, note 1) [92].
[31] *Google Spain* (supra, note 1) [81].
[32] *Google Spain* (supra, note 1) [99].
[33] See for instance: E Frantziou, 'Further Developments in the Right to be Forgotten: The European Court of Justice's Judgment in Case C-131/12, Google Spain, SL, Google Inc v Agencia Espanola de Proteccion de Datos' (2014) 14 Human Rights Law Review 761, 769; C Kuner, 'The Court of Justice of the EU Judgment on Data Protection and Internet Search Engines' LSE Law, Society and Economy Working Papers 3/2015 <http://papers.ssrn.com/sol3/papers.cfm?abstract_id=2496060> last accessed

mention the right to impart and receive information of search engine operators and of people who publish information on the web.[34] The CJEU refers to the right to receive information of searchers only as 'interests', without mentioning the fundamental right to receive information.[35]

We are not suggesting that the CJEU should have interpreted the directive in such a way that Costeja Gonzáles, who sought to delist an out-dated newspaper article about his past debts from Google search results, would have lost his case. Rather, we fear that the Google Spain judgment could negatively impact other decisions on delisting requests by search engine operators, data protection authorities, and national courts. At a minimum, the CJEU should have explicitly considered the search engine operator's right to freedom of expression and information, and should have given more attention to people's right to receive and impart information. The CJEU suggests that 'as a rule', privacy and data protection rights override the public's interest in finding information.[36] We fear that search engine operators, data protection authorities, and national courts might therefore not adequately consider the right to freedom of expression in their delisting decisions based on *Google Spain*.

The CJEU's remark that privacy and data protection rights override 'as a rule' the right to information is hard to reconcile with the case law of the European Court of Human Rights.[37] The latter Court says privacy and freedom of expression rights have equal weight.

---

on 6 August 2015; JVJ Van Hoboken, 'The Google Spain/Costeja Decision', European Human Rights Cases 2014/186, 2014.
[34] Kulk and Zuiderveen Borgesius (supra, note 2).
[35] See also: O Lynskey 'Control over Personal Data in a Digital Age: Google Spain' (2015) 78(3) MLR 522, 531.
[36] See Section II.1.
[37] See article 52(3) of the Charter of Fundamental Rights of the European Union: "In so far as this Charter contains rights which correspond to rights guaranteed by the Convention for the Protection of Human Rights and Fundamental Freedoms, the meaning and scope of those rights shall be the same as those laid down by the said Convention. This provision shall not prevent Union law providing more extensive protection." See also: P Lemmens, 'The Relation between the Charter of Fundamental Rights of the European Union and the European Convention on Human Rights-Substantive Aspects' (2001) 8 Maastricht Journal of European and Comparative Law 49.

The Article 29 Working Party, in which national Data Protection Authorities cooperate, has published guidelines for implementing the *Google Spain* judgment.[38] Unlike the CJEU, the Working Party does mention the right to receive and impart information, and says search engine operators should take that right into account when deciding about delisting requests.[39]

Although the right to be delisted was recognised by the CJEU, decisions about actual delisting requests need to be decided on by national courts and data protection authorities.[40] National courts and data protection authorities must consider the CJEU's guidance when taking their decisions. In the next section we explore the application of the *Google Spain* judgment in the Netherlands.

## III  Delisting requests in the Netherlands

### III.1  Non-litigated delisting requests

In the Netherlands, Google's search engine has a market share of around 85%.[41] As the two Dutch court cases involved Google, and the Dutch Data Protection Authority only mentioned delisting requests regarding Google, we focus our further analysis on Google.[42]

Since *Google Spain*, Google receives many delisting requests.[43] At Google, between fifty and one hundred people are working fulltime on dealing with worldwide delisting requests.[44] Of the requests received, over 17,000 came from the Netherlands.[45] Google removed roughly 42% of the requested URLs for search queries for people's names.[46] Google published some examples of delisting requests it received, and one of those is from the Netherlands. In this example somebody asked Google 'to remove over 50 links to articles and blog posts reporting on public outcry over accusations that he was abusing welfare services.'[47] Google refused to delist the links.[48]

---

[38] Article 29 Data Protection Working Party 'Guidelines on the Implementation of the Court of Justice of the European Union Judgment on "Google Spain and Inc v. Agencia Española de Protección de Datos (AEPD) and Mario Costeja González" C-131/12' (2014) WP 225.
[39] ibid, 6.
[40] True, many delisting requests are decided on by search engine providers without scrutiny by the courts. We return to that point in section IV.1.
[41] iProspect 'Nationale Search Engine Monitor Onderzoek' (2014) <www.iprospect.com/nl/nl/press-room/nationale-search-engine-monitor-onderzoek> last accessed on 6 August 2015.
[42] See also Article 29 Working Party, press release 18 June 2015 <http://ec.europa.eu/justice/data-protection/article-29/press-material/press-release/art29_press_material/20150618_wp29_press_release_on_delisting.pdf> last accessed on 6 August 2015: "The majority of complaints concerned Google Inc.'s search engine."
[43] As of August 3 2015, Google has received more than 290,000 requests in total. Google 'Transparency Report. European privacy requests for search removals' <www.google.com/transparencyreport/removals/europeprivacy> last accessed on 6 August 2015.

The Dutch Data Protection Authority (*College Bescherming Persoonsgegevens*) refers on its website to the delisting guidelines by the Article 29 Working Party.[49] In November 2014, the Dutch Data Protection Authority did report that it had received complaints from over 30 people whose delisting requests had been rejected by Google.[50] In two cases the Authority asked Google to delist search results with which Google complied. In three other cases the Authority asked Google to reconsider its refusal to delist search results; Google promised to do so.

In all other cases, the Data Protection Authority refrained from mediation and told the complainants that they should go to court. Most of the rejected complaints came from people with a role in public life, such as former politicians and executives, and professionals working in financial industries and healthcare. The Authority rejected other complaints because it was unable to assess whether the information in question was inaccurate or out-dated. Finally, the Authority noted that it does not have the competence to rule on requests to remove defamatory or libellous statements.[51] Since

---

[44] As reported by Peter Fleischer, Google's Global Privacy Counsel, at the Privacy & Innovation Conference at Hong Kong University, 8 June 2015, <www.lawtech.hk/pni/?page_id=11> last accessed on 6 August 2015.
[45] Google 'Transparency Report. European privacy requests for search removals' <www.google.com/transparencyreport/removals/europeprivacy> last accessed on 6 August 2015.
[46] ibid.
[47] ibid.
[48] ibid.
[49] The Dutch Authority has not developed, at least has not published, its own guidelines regarding delisting requests.
[50] College Bescherming Persoonsgegevens, 'Behandeling afgewezen verwijderverzoeken Google' <https://cbpweb.nl/nl/nieuws/behandeling-afgewezen-verwijderverzoeken-google> last accessed on 6 August 2015.
[51] See article 3 of the Dutch Data Protection Act: "1.This Act does not apply to the processing of personal data for exclusively journalistic, artistic or literary purposes, except where otherwise provided

*Google Spain*, two cases about delisting requests went to court in the Netherlands; we turn to those cases now.

### III.2 Convicted criminal court case

**III.2.a Background**

In 2012, a popular commercial Dutch TV channel aired a programme of the famous Dutch crime news reporter Peter R. de Vries. The programme featured hidden camera footage of a man discussing with an assassin how best to kill a competitor. The TV programme did not refer to the man's full name, but only to his first name and the first letter of his last name ('initials'): Arthur van M. In the Netherlands, the media usually refer to people involved in criminal proceedings this way.[52]

The TV programme's footage was used as evidence in a criminal case against Arthur van M. In 2012 he was convicted and sentenced to six years imprisonment for attempting to incite an assassination. Arthur van M appealed and is awaiting judgment in that criminal case.

The conviction and the TV show were widely reported. The media reports only mentioned Arthur van M's initials. The case inspired Antoon Engelbertink to write the 2013 book, *The Amsterdam Escort-assassination*. An English translation was also published. Engelbertink described his book as a mix between fact and fiction. In the book, a man who commissioned an assassination has the same name as Arthur van M.

If a searcher entered the full name of Arthur van M into Google, the search engine displayed a set of URLs. Arthur van M wanted to have some of these URLs delisted. Some of the URLs referred to web pages about Engelbertink's book on amazon.com, books.google.nl, and abebooks.com. For searches on Arthur van M's full name,

---

in this Chapter and in Articles 6 to 11, 13 to 15, 25 and 49" (unofficial translation at <www.coe.int/t/dghl/standardsetting/dataprotection/National%20laws/NL_DP_LAW.pdf> last accessed on 6 August 2015).
[52] For more details see: Netherlands Press Council 'Guidelines'
<www.rvdj.nl/uploads/fckconnector/192f9e9a-ece2-4f50-9f59-2952d7835de3> last accessed on 6 August 2015.

Google's autocomplete function also suggested to add 'peter r de vries' (the crime reporter's name) to the queries. Search results pages also displayed a message that 'Some results may have been removed under data protection law in Europe'.

Google refused to grant Arthur van M's requests to remove these URLs from searches for his name. At the District Court of Amsterdam, Arthur van M invoked *Google Spain* and claimed, in short, that Google should delist certain search results relating to him. In total, Arthur van M submitted five claims; the District Court rejected them all.[53] He therefore appealed the decision to the Court of Appeal of Amsterdam. In the next section, we discuss his claims and the Amsterdam Court of Appeal's decision.

**III.2.b Court decision**

The Court of Appeal notes that Arthur van M is being prosecuted for a serious criminal offence for which he has been convicted in first instance.[54] The Court assumes that any publications about his conviction are the result of the public's interest in convictions, and his own unlawful behaviour. The public is very interested in reading about serious criminal offences in general, and thus also about the prosecution and conviction of Arthur van M.

Furthermore, the Court notes that search results for Arthur van M's full name point only to websites that refer to his first name and the first letter of his last name ('initials'). Therefore, people looking for information about him who already know his full name, cannot be sure whether the information on the websites refers to him or to somebody else, because the web pages only contain the initials. The Court says that, in some circumstances, searchers might make a connection between the website's content and Arthur van M, for example because they know about the activities of his escort agency, or know other information that identifies him as the crime suspect. But because of the role Arthur van M plays in public life, and the crime he committed, he must accept people can make such a connection.

---

[53] Rechtbank Amsterdam, 18 September 2014, ECLI:NL:RBAMS:2014:6118.
[54] Gerechtshof Amsterdam, 31 March 2015, ECLI:NL:GHAMS:2015:1123.

Arthur van M argued that searchers could use the book, in which his full name is used, to connect him to the criminal offence. The Court, however, notes that the book is presented as a mix between fact and fiction. In addition, the book describes an actual assassination, instead of an attempt to arrange an assassination. Thus, on the basis of the book or references to it, the public cannot connect Arthur van M with the crime. If however the public does make such a connection, the Court finds that Arthur van M has to live with that.

The first claim of Arthur van M was as follows. He claimed that Google, under penalty of a fine, should be required to: 'correct, delete, and/or block his personal data by removal of the URLs which relate to him when his name is entered into Google's search engine.'[55] Because Arthur van M is prosecuted for a recently committed offence and was convicted at first instance, he does not have the right to have search results removed that might link him to the offence. The Court concludes that the URLs should not be delisted.

Second, Arthur van M wanted Google to be required to 'remove, and keep removed, all search results that refer, or referred, to his personal data.'[56] The Court, however, says that the search results only refer to the book or a website containing Arthur van M's initials. Because there is no clear connection between the book and his identity, the Court does not order Google to delist search results that only refer to the book or his initials. Additionally, the Court notes that, while it is common for Dutch media to refer to people involved in criminal proceedings by their initials, there is no enforceable rule to that effect. The Court says Arthur van M's claim to 'keep removed' certain search results is formulated too broadly and too imprecise, and for that reason alone cannot be granted.

Third, Arthur van M asked for removal of the message that some search results may have been removed under European data protection law, for searches for his name. The Court accepts Google's argument that the message is a standard message for name searches, and that this message is independent from any delisting requests. Therefore, the Court rejects Arthur van M's third claim.

---

[55] Our translation. Gerechtshof Amsterdam, 31 March 2015, ECLI:NL:GHAMS:2015:1123 [3.2].
[56] Our translation. Gerechtshof Amsterdam, 31 March 2015, ECLI:NL:GHAMS:2015:1123 [3.2].

Fourth, Arthur van M wanted Google to 'remove the connection between his name and the name of crime reporter "Peter R. de Vries" in Google's search bar.'[57] Arthur van M argued that Google aimed to discredit him, by having the autosuggest function suggesting the name of crime reporter Peter R. de Vries for searches on his full name. But the Court concluded that Arthur van M had not convincingly shown that Google aimed to harm him. Moreover, the public has a legitimate interest in being informed about Arthur van M and his crimes. Finally, the Court argues that users who are confronted with the autosuggestion of the crime reporter's name apparently already know the full name of Arthur van M, and will not find more information through search results generated by the autosuggest addition.

Fifth, the man essentially asked the Court to order Google to never breach his privacy again.[58] The Court rejects that claim because it is too broad. In sum, all of Arthur van M's claims were rejected: he lost the case.

**III.3    KPMG partner court case**

**III.3.a Background**

In 2011, a partner at KPMG, a large firm providing audit, tax and advisory services, commissioned a contractor to build a new house. During construction, the KPMG partner and his family lived in a house next door. Three portakabin containers were also installed to provide extra space.

The KPMG partner became engaged in a dispute with the contractor. The contractor said that the KPMG partner still had to pay 200,000 euros due to additional work and late payments. Therefore the contractor replaced the door locks so the KPMG partner could not enter the new house. The KPMG partner and the contractor brought their dispute before the Dutch Arbitration Board for the Building Industry. They settled their dispute. As part of the settlement, the KPMG partner paid the contractor around 60,000 euros.

---

[57] Our translation. Gerechtshof Amsterdam, 31 March 2015, ECLI:NL:GHAMS:2015:1123 [3.2].
[58] Arthur van M wanted Google to be ordered to: 'refrain from any infringements on his right to privacy by making available or reproducing the URLs in Google's search engine in relation to automated processing of his name for commercial purposes, or other similar public and/or commercial communication, in any form or way, including in Google Books' (our translation). Gerechtshof Amsterdam, 31 March 2015, ECLI:NL:GHAMS:2015:1123 [3.2]). Also in Dutch, the claim is difficult to read.

The largest Dutch newspaper, *De Telegraaf,* published a front-page article on the matter in 2012: 'KPMG Top Executive Camps in Container'.[59] The *Telegraaf* article reports that the KPMG partner cannot move into his new villa because his contractor changed the locks in revenge for an unpaid bill. The article says the KPMG partner and his wife disliked the paintwork, and wanted the contractor to pay damages for the emotional harm they suffered because they had to live in containers longer than planned. The story was also reported on other news websites.

In 2014, the KPMG partner requested Google to remove the *Telegraaf* article from search results for searches for his name. Google refused. Later, the KPMG partner's lawyer requested the removal of several URLs for searches relating to the KPMG partner. Google again rejected the request, and said the webpages contained information that is relevant, of public interest, and not out-dated.

The KPMG partner filed a case with the District Court of Amsterdam. Invoking *Google Spain*, he wanted Google to be ordered to delist the search results for the URLs, and any other webpages associating him with the container story. In case his first claim failed, the KPMG partner asked the Court to order Google to move the search results that refer to him to the bottom of all search results.

---

[59] An online version of the *Telegraaf* article is still available on the web. The article mentions the full name of the KPMG partner and is accompanied by a photo of him. While the Court's decision has been anonymised and does not mention the KPMG partner's name, the Court refers to the article's URL in its decision effectively neutralizing the partner's anonymity: <www.telegraaf.nl/binnenland/20051811/__Topman_KPMG_in_container__.html> last accessed on 6 August 2015. De Telegraaf has been described as "a conservative-oriented daily with populist tendencies" (M Hajer and W Versteeg, 'Political rhetoric in the Netherlands: reframing crises in the media' (2009) 7 footnote 20). See also: D Trilling and K Schoenbach, 'Investigating people's news diets: How online news users use offline news' (2015) 40(1) Communications: The European Journal of Communication Research 67-91, 75.

The KPMG partner argued that the contractor acted unlawfully by changing the locks. The KPMG partner said that the contractor did not have a valid claim of 200,000 euros. According to the KPMG partner, The *Telegraaf* article harms him because many people he meets, for instance when his children join a new hockey team, refer to the story. Furthermore, the article harms the KPMG partner's career, as new clients often search for him on Google and find the container story. He anticipates that the search results will also harm him in future job applications. According to the KPMG partner, the container story is irrelevant for the general public, because it concerns a private issue that had nothing to do with his position at KPMG. He adds that the information is irrelevant because it is two-and-a-half years old.

**III.3.b Court decision**

In contrast to the CJEU in *Google Spain*, the Dutch court starts with emphasising that search engines, such as Google, play an important role in society.[60] The Court says the internet contains 'an ocean of information', which, moreover, may change at any moment. Search engines help people to find information online. If search engines were subject to too many restrictions, their cataloguing function would be hampered, resulting in a loss of credibility for those search engines.

The Court says two fundamental rights are stake. Firstly, the KPMG partner's right to privacy as protected by the European Convention on Human Rights.[61] Secondly, Google's right to 'freedom of information'[62] (as the Court calls the right to receive and impart information), protected by the Convention and the Dutch Constitution.[63] The Court adds that the interests of internet users, webmasters, and authors of online information should be taken into account as well.

The Court stresses that the relevance of the search results is at issue – not the relevance of the press publications. If the KPMG partner seeks a review of the content of a publication-he should sue the original publisher. A court can then apply the legal framework regarding unlawful press publications, and assess whether a particular publication is sufficiently supported by evidence. However, people should not be able to circumvent the legal framework for unlawful press publications by invoking Articles 12 and 14 of the Data Protection Directive. The Court adds that these provisions are not meant to help people hiding lawful publications from the public through a removal request, merely because they dislike those publications.

---

[60] Rechtbank Amsterdam, 13 February 2015, ECLI:NL:RBAMS:2015:716. See also: case note by L. Gorzeman & P. Korenhof (in Dutch), for Rechtbank Amsterdam, 13 February 2015, ECLI:NL:RBAMS:2015:716, in Computerrecht 2015/86; P. Kreijger, 'Een jaar later: de receptie van 'het recht vergeten te worden' in de Nederlandse rechtspraak', Mediaforum 2015-4, p. 141-148.
[61] See art. 8 of the European Convention on Human Rights.
[62] Our translation. Rechtbank Amsterdam, 13 February 2015, ECLI:NL:RBAMS:2015:716 [4.4].
[63] See art. 10 of the European Convention on Human Rights; art 10 of the Dutch Constitution.

Regarding the right to request erasure of personal data (Article 12(b) of the Data Protection Directive), the Court notes that the URLs are accurate search results for a search for the KPMG partner's name, because the information on the websites behind those URLs relates to him. In addition, the information provided on the URLs is essentially correct. Google argued that the publications were written when KPMG was involved in several financial scandals. Furthermore, during that period there was much discussion about financial morals of top business executives such as the KPMG partner. Google added that prominent national and local media published the stories. Apparently these media considered the information newsworthy – an important factor for Google when determining the relevance of information.

The Court concludes that it is plausible that the search results in question are relevant and not excessive. Therefore, the Court rejects the KPMG partner's erasure request (Article 12(b) of the directive). Regarding the question whether the information is still up-to-date, the Court holds that the KPMG partner's case is not (yet) comparable to *Google Spain*, which concerned a 16-year-old publication.

Regarding the right to object to processing (Article 14(a) of the directive), the Court holds that there are no compelling grounds relating to the KPMG partner's situation that prohibit processing his personal data. According to the Court, the 'right to deletion' is an exception to Google's right to freedom of information, and should therefore not be granted easily. The Court understands that the KPMG partner finds it unpleasant to be confronted repeatedly with the container story by acquaintances and business contacts. Yet, this displeasure does not outweigh Google's right to freedom of information. The Court adds that the publications are not defamatory. While the publications suggest that the KPMG partner had a dispute with the contractor, they do not suggest that he was to blame. Furthermore, according to the Court the statement that the KPMG partner had to reside in temporary housing longer than planned is not defamatory. After all, the KPMG partner had lived in temporary housing for months. Therefore, the Court refuses the KPMG partner's claim regarding delisting search results.

As noted, the KPMG partner had a subsidiary claim: he wanted Google to move certain search results to the bottom of all search results. The Court notes that Google said it is technically impossible to influence the search results in such a way that a particular URL is listed on a particular search engine result page. The Court, apparently accepting Google's argument, rejects the KPMG partner's claims. (It seems questionable whether it is really impossible for Google to demote search results. Perhaps the Court accepted Google's statement as true because the man did not sufficiently debunk Google's argument during the proceedings.)

The Court focuses its analysis on the lawfulness of the search results – not on the lawfulness of the publication that the search results refer to. The Court draws a line: if the publication's content is the problem, the data subject should address the publisher – not the search engine operator. In this context, the Court also notes that the right to be delisted should not be used to circumvent proceedings against the publisher of the content to hide lawful but inconvenient publications from the public. The Dutch court takes a different approach than CJEU, as the CJEU in *Google Spain* made clear that even if the publication itself is lawful, results may have to be delisted.[64]

## IV    Consideration of freedom of expression in the Netherlands

### IV.1    Non-litigated delisting requests

To assess the effects of the *Google Spain* judgment in the Netherlands, two separate conclusions must be drawn, regarding delisting requests that were litigated in court, and delisting requests that never made it to court. We first take a look at the non-litigated requests.

---

[64] *Google Spain* (supra, note 1) [88].

We do not know how many cases Google decided in favour of requesters in cases that did not go to court, in which the request should have been denied because more weight should have been given to freedom of expression. Google's opaque decision procedures make it difficult to assess the implications for freedom of expression that follow from the *Google Spain* judgment.

Little information is available about how Google has assessed the 17,000 delisting requests it received from the Netherlands. Google reported that it delisted URLs in about 42% of all Dutch cases.[65] However, Google shares little about the nature of those delisted URLs. Were they news articles, Wikipedia pages, or information in public records? Do requests come from laypeople, politicians, or criminals? The Guardian reports that 98% of the delisting requests to Google came from ordinary citizens.[66] This percentage seems to suggest that the right to delist information satisfies a need among ordinary people to protect their privacy.

Because of this lack of transparency, an international group of academics has called upon Google to provide more information about its decision procedure regarding delisting requests. 'Beyond anecdote, we know very little about what kind and quantity of information is being delisted from search results, what sources are being delisted and on what scale, what kinds of requests fail and in what proportion, and what are Google's guidelines in striking the balance between individual privacy and freedom of expression interests.'[67]

Furthermore, giving private parties, such as search engine operators, the task to balance fundamental rights has its drawbacks. There may be clear-cut cases where privacy should prevail over freedom of expression, and in which it makes sense that a search engine operator delists a search result. However, in more difficult cases, search engine operators may not be the most appropriate party to balance the fundamental rights involved. The case of Arthur van M illustrates that deciding about a delisting request can be difficult. For instance, the fact that the information on the web did not directly refer to him, but instead referred to information in a fact-fiction book in which one of the characters bears the same name as Arthur van M, raises the question if personal data were involved at all.

---

[65] Google 'Transparency Report. European privacy requests for search removals' <www.google.com/transparencyreport/removals/europeprivacy> last accessed on 6 August 2015.
[66] S Tippmann and J Powles, 'Google accidentally reveals data on 'right to be forgotten' requests',<www.theguardian.com/technology/2015/jul/14/google-accidentally-reveals-right-to-be-forgotten-requests> last accessed on 6 August 2015.
[67] Open Letter to Google From 80 Internet Scholars: Release RTBF Compliance Data <https://medium.com/@ellgood/open-letter-to-google-from-80-internet-scholars-release-rtbf-compliance-data-cbfc6d59f1bd> last accessed on 6 August 2015.

In sum, it is difficult to assess the implications for freedom of expression that follow from *Google Spain*, because of the lack of information about Google's decisions.

**IV.2    Litigated delisting requests**

So far, in delisting cases, Dutch courts give more attention to freedom of speech than the CJEU did in *Google Spain*. The Dutch courts emphasise the important societal role of search engines in finding information. In the KPMG partner case, the District Court of Amsterdam explicitly considers the search engine operator's right to freedom of information. Furthermore, the Dutch court stresses the decisive role that search engines play in facilitating access to information. In contrast, the CJEU focused primarily on the role of search engines in disseminating personal data.

If, in the KPMG partner case, the Dutch court had more closely followed the CJEU's *Google Spain* reasoning, the outcome would probably have remained the same for the KPMG partner. Arguably the KPMG partner played a role in public life – the criterion used by the CJEU. Under *Google Spain*, and under case law of the European Court of Human Rights, people who play a role in public life must accept more privacy interferences than less well-known people. However, in contrast to the CJEU, the Dutch court explicitly considers the right to freedom of expression. While the Dutch court's decision to reject the delisting request seems compatible with *Google Spain*, the Dutch court recognises that the right to privacy and freedom of expression deserve equal weight.

In the case regarding the convicted criminal, the District Court of Amsterdam mentions Google's important information-retrieval function, and Google's right to freedom of expression. The Court of Appeal confirms the District Court's decision, but closely follows the CJEU's line of reasoning that focuses mainly on the data subject's privacy and data protection rights. Nevertheless, the Court of Appeal arrives at a decision that seems reasonable from a freedom of expression perspective.

The claimants in the Dutch cases did not have a case as strong as Costeja Gonzáles did. Both cases concerned information that was more recent than the financial problems of Costeja Gonzáles. Furthermore, both Dutch cases concern claimants with a higher public profile than Costeja Gonzáles.

For instance, if Arthur van M is indeed guilty, he took the risk that media would report about him. As the District Court of Amsterdam notes, 'the right to privacy is not absolute.'[68] Furthermore, 'committing a crime leads to being in the news in a very negative way; this leaves traces on the internet – possibly even for a very long time.'[69]

This line of reasoning fits with the approach of the European Court of Human Rights, which holds that 'the public do, in principle, have an interest in being informed – and in being able to inform themselves – about criminal proceedings.'[70] The European Court of Human Rights also deems the seriousness of the crime relevant.[71] The Dutch court's finding that the public has great interest in reporting on serious crimes accords with the approach of the European Court of Human Rights. The Dutch court might have decided differently if the case concerned an ex-criminal trying to build a new life after twenty years in jail. However, the Arthur van M case concerns a crime of a few years ago.

The KPMG partner's argument that his dispute with the contractor concerned a private matter has some merit. But the dispute happened during a time when the morals of top executives were much debated. Furthermore, it could be argued that top executives are, because of their job choice, public figures. Moreover, the Dutch Public Prosecution Service investigated KPMG and three former partners for covering up bribes paid in Saudi Arabia by a large construction and engineering company.[72] Hence, KPMG and its partners were indeed a topic of public debate.[73] On the other hand, it could also be argued that the KPMG partner's dispute with the contractor did not have anything to do with his job. In sum, it's not straightforward whether articles regarding the KPMG partner's dispute contribute to the public debate. But it does seem reasonable to argue that the articles did.

---

[68] Rechtbank Amsterdam, 18 September 2014, ECLI:NL:RBAMS:2014;6118 [4.20].
[69] Rechtbank Amsterdam, 18 September 2014, ECLI:NL:RBAMS:2014;6118 [4.7].
[70] *Axel Springer AG v Germany* App no 39954/08 (ECtHR 7 February 2012) [96].
[71] *Axel Springer AG v Germany* App no 39954/08 (ECtHR 7 February 2012) [100].

Overall, the Dutch courts' approach of explicitly taking into account the right to freedom of expression differs from the CJEU's approach. But if, in a case comparable to the one of Costeja Gonzáles, somebody requested Google to delist pages about old and irrelevant debts for searches his or her name, it is plausible that a Dutch court would decide in favour of the requester.

Indeed, our point is not that as many delisting requests as possible should be denied for the sake of free expression. Rather, our point is that courts should properly balance freedom of expression, and privacy and data protection. This balancing should happen on a case-by-case basis, taking all relevant circumstances into account. As noted, the European Court of Human Rights has developed nuanced and detailed case law, which recognises the need to balance the rights of freedom of expression and privacy when they are in conflict. Data protection law does recognise the importance of free expression, for instance through the media exception.[74] Nevertheless the framework developed by the European Court of Human Rights allows for a more nuanced balancing act between freedom of expression and privacy.

---

[72] Openbaar Ministerie, 'KPMG treft schikking voor haar rol bij het verhullen van betalingen aan buitenlandse agenten' <www.om.nl/actueel/nieuwsberichten/@32396/kpmg-treft-schikking> last accessed on 6 August 2015.

[73] The European Court of Human Rights has said in a case where a newspaper reported on a fraud case, mentioning a bank manager by name: "there is little scope under Article 10 § 2 of the Convention for restrictions on political speech or on debate on questions of public interest" (*Standard Verlags Gmbh v. Austria* App no 34702/07 (ECtHR 10 January 2012), [40]).

[74] See Erdos (supra, note 25).

# V    Conclusion

Focusing on freedom of expression, this paper examined how the *Google Spain* judgment has been applied in the Netherlands. Since the *Google Spain* judgment of the CJEU, people have, under certain conditions, the right to have search results for their name delisted. It is unknown how Google decided on the 17,000 delisting requests it received in the Netherlands. The opaqueness of Google's decision procedures makes it difficult to assess the effect of *Google Spain* on freedom of expression. This lack of transparency points to a general problem with giving private parties, such as search engine operators, the task to balance fundamental rights.

In *Google Spain*, the CJEU suggested that privacy and data protection rights generally override the public's right to receive information. Since *Google Spain*, Dutch courts have decided on two cases regarding delisting requests. So far, compared to the CJEU, Dutch courts seem more conscious of the need to balance privacy and data protection rights on the one hand, and the right to freedom of expression on the other.

* * *